\documentclass[aps,prl,preprint,superscriptaddress]{revtex4-1} 

 \usepackage[utf8]{inputenc}
\usepackage{amssymb}
 \usepackage{amsmath,bm}
 \usepackage{mathrsfs}
 \usepackage{amsfonts}
 \usepackage{graphicx}
 \usepackage{setspace} 
 \usepackage{graphicx}
 \usepackage{epstopdf}
 \usepackage{dcolumn}
 \usepackage{amsmath}
 \usepackage{epsfig}
 \usepackage{indentfirst}
 \usepackage{psfrag}
 \usepackage{subfigure}
 \usepackage{amssymb}
 \usepackage{color}
 \usepackage{xcolor}
 \usepackage{siunitx}
\usepackage{xspace}
 \usepackage{graphicx}
 \usepackage{dcolumn}
 \usepackage{bm}
 \usepackage{natbib}
\usepackage{physics}
\usepackage[modulo]{lineno}
\modulolinenumbers[5]

\usepackage[backref=none,bookmarksnumbered=true,bookmarks=true,bookmarksopen=true,colorlinks=true,
citecolor=blue,linkcolor=blue,anchorcolor=green,urlcolor=blue,unicode=false]{hyperref}

\usepackage{ulem}[normalem]

\def\spin12{spin-$\tfrac{1}{2}$}
\def\didv{$dI/dV$\xspace}

\makeatletter

\newcommand\colorsout[1]{\bgroup \markoverwith{\textcolor{#1}{\rule[0.5ex]{2pt}{0.4pt}}}\ULon}

\makeatother

\begin{document}

\title{Parity-dependent coupling of  molecular spin chains \\to a superconductor}

\setstretch{1.1} 
\author{Katerina Vaxevani}\thanks{These authors contributed equally to this work}
    \affiliation{CIC nanoGUNE-BRTA, 20018 Donostia-San Sebasti\'an, Spain}

\author{Jon Ortuzar}\thanks{These authors contributed equally to this work}
      \affiliation{CIC nanoGUNE-BRTA, 20018 Donostia-San Sebasti\'an, Spain}
  \affiliation{Quantronics group, Service de Physique de l'\'Etat Condens\'e \mbox{(CNRS, UMR 3680)}, IRAMIS, CEA-Saclay, Universit\'e Paris-Saclay, 91191 Gif-sur-Yvette, France}
  
\author{Stefano Trivini}\thanks{These authors contributed equally to this work}
    \affiliation{CIC nanoGUNE-BRTA, 20018 Donostia-San Sebasti\'an, Spain}

\author{Georg Monninger}
    \affiliation{CIC nanoGUNE-BRTA, 20018 Donostia-San Sebasti\'an, Spain}

\author{Dongfei Wang}
    \affiliation{CIC nanoGUNE-BRTA, 20018 Donostia-San Sebasti\'an, Spain}
    \affiliation{School of Physics, Beijing Institute of Technology, Beijing 100081, China}
 
\author{Manuel Vilas-Varela}
    \affiliation{Centro Singular de Investigaci\'on en Qu\'imica Biol\'oxica e Materiais Moleculares (CiQUS) and Departamento de Qu\'imica Org\'anica, Universidad de Santiago de Compostela, 15782 Santiago de Compostela, Spain}
    
\author{Lucía Gómez-Rodrigo}
    \affiliation{Centro Singular de Investigaci\'on en Qu\'imica Biol\'oxica e Materiais Moleculares (CiQUS) and Departamento de Qu\'imica Org\'anica, Universidad de Santiago de Compostela, 15782 Santiago de Compostela, Spain}

\author{Diego Peña}
    \affiliation{Centro Singular de Investigaci\'on en Qu\'imica Biol\'oxica e Materiais Moleculares (CiQUS) and Departamento de Qu\'imica Org\'anica, Universidad de Santiago de Compostela, 15782 Santiago de Compostela, Spain}
    \affiliation{Oportunius, Galician Innovation Agency (GAIN), 15702 Santiago de Compostela, Spain}  

\author{Jose Ignacio Pascual} \email{ji.pascual@nanogune.eu}
    \affiliation{CIC nanoGUNE-BRTA, 20018 Donostia-San Sebasti\'an, Spain}
    \affiliation{Ikerbasque, Basque Foundation for Science, 48013 Bilbao, Spain}

\begin{abstract}
Topological order can fractionalize the quantum numbers of the underlying particles. A paradigmatic example is the spin-1/2 states at the edges of an antiferromagnetic  integer-spin chain, protected by a topological Haldane gap in the bulk and mutually coupled in short chains. Owing to their topological protection, they are natural building blocks for hybrid spin–superconductor quantum systems. Whether fractionalization survives the coupling to a superconducting condensate, however, remains an open question. Here we grow molecular Haldane chains of antiferromagnetically coupled spin-1 triangulene units on a proximitized Au(111)/Nb(110) surface and resolve a parity-dependent coupling of their spin-1/2 edge states to the superconducting condensate by scanning tunnelling spectroscopy. Odd-length chains host Yu–Shiba–Rusinov bound states inside the superconducting gap, originating from the net S=1 ground state, whereas even-length chains form an S=0 ground state decoupled from the superconductor. A two-site superconductor model reveals that this alternation arises from the sign and strength of the inter-edge interaction, a mechanism independently validated by extra-gap spin excitations in tunnelling spectra. Collective many-body spin excitation modes are also detected decoupled from the superconductor by the much larger Haldane gap. The length-tunable coupling of the edge spins to the superconductor opens a route toward molecular spin qubits based on $\pi$-conjugated carbon architectures. 
\end{abstract}
\date{\today}

\maketitle

\begin{figure*}[b]
        \begin{center}
			\includegraphics[width=\textwidth]{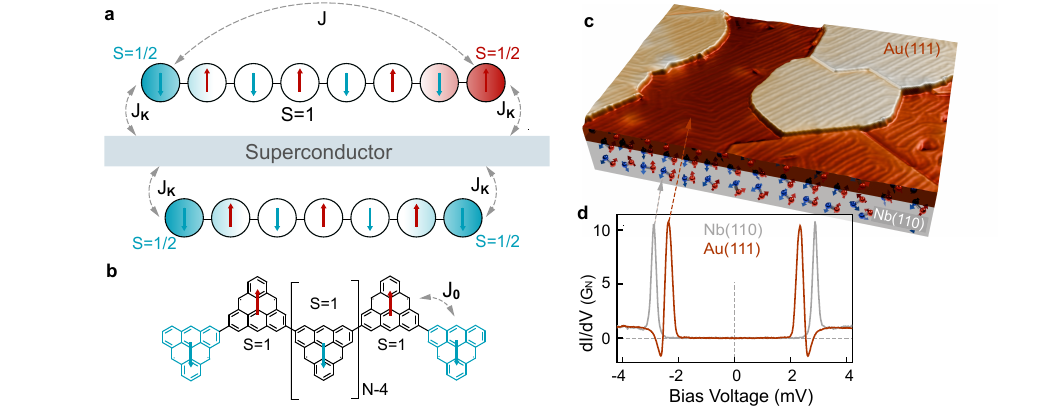}    
        \end{center}
			\caption{ \textbf{Interacting spin chains and a proximitized Au(111)/Nb surface:} \textbf{a} Representation of an S=1 antiferromagnetic chain with fractional spin-1/2 edge states. Depending on the parity of the chain length \textit{N}, the ground state is S=0 (even) or S=1 (odd), set by the relative orientation of the edge spins. \textbf{b} Chemical structure of a S=1 triangulene chain of length N used in this work as integer spin chain \cite{Mishra2021}. \textbf{c} STM topography of a Au film   grown on Nb(110), exhibiting the characteristic herringbone reconstruction of Au(111) surfaces. \textbf{d} Differential conductance spectrum measured with a Nb tip on the clean Au(111)/Nb(110) surface (red), compared with a spectrum on the bare Nb(110) substrate (grey), with sharp quasiparticle peaks from de Gennes-Saint James resonances \cite{deGennes1963} and absolute gap of size $\Delta_{\mathrm{dGSJ}}<\Delta_{\mathrm{Nb}}$ (\didv @ V = 5 mV, I = 500 pA).             }
		\label{fig1}
		\end{figure*}

\setstretch{1.4} 
\textit{Introduction:} 
Exotic phases of matter often emerge when collective interactions reshape the fundamental particles into fractional entities with new quantum numbers. Such fractionalization is a hallmark of strongly correlated systems, manifesting in phenomena ranging from fractionally charged quasiparticles in the quantum Hall effect to solitons in polymers and Majorana modes in topological superconductors \cite{Su1979,Jackiw1976,Kitaev2001}. In one-dimensional antiferromagnetic integer spin chains, fractionalization gives rise to symmetry-protected spin-1/2 edge states \cite{Haldane1983PRL}, a paradigmatic instance of topological order~\cite{Tasaki2025}. As topologically protected two-level systems, end states are considered building blocks for quantum spin devices \cite{Miyake2010,jaworowski2017}. In finite chains,  their mutual interaction yields parity-dependent singlet or triplet ground states (Fig.~\ref{fig1}a), an entangled manifold attractive for quantum information schemes \cite{Brennen2008,Miyake2010,Castillo2026}. Their integration on a superconducting condensate is a key step toward hybrid quantum devices that exploit both topological spin correlations and superconducting pairing~\cite{Huang2026,Baran2024,tenhaaf2025}. 

One possible route for this integration is the bottom-up synthesis ~\cite{Clair2019} of atomically precise molecular nanostructures  on superconducting surfaces. It has been shown that these molecular platforms can recreate paradigmatic correlated systems and symmetry-protected topological order \cite{Cao2017b,Rizzo2018,Groning2018,Cirera2020,Li2021,Mishra2021,Zhao2024b,Zhao2025b,Sun2025,Su2025,Yuan2025}. In particular, triangulene integer-spin chains (Fig.~\ref{fig1}b) synthesized on Au(111) exhibit the hallmarks of the Haldane phase, namely gapped bulk excitations and fractional $S=1/2$ edge states~\cite{Mishra2021}.  Thus, their synthesis on superconducting substrates would offer an ideal environment to probe the coupling of their edge states to Cooper pairs and their excitations: the quasiparticle gap protects spin excitations from rapid damping ~\cite{Heinrich13a,Vaxevani2022extending}, while Yu–Shiba–Rusinov (YSR) bound states provide a spectroscopic fingerprint of localized magnetic moments ~\cite{yu1965bound,shiba1968classical,rusinov1969,Heinrich2018}. Yet integrating molecular spin chains with superconductors has remained challenging ~\cite{liu_proximity-induced_2023,Wang_epitaxial_2025} because their on-surface synthesis requires the catalytic properties of bulk gold, incompatible with conventional superconducting surfaces. A platform that preserves this activity while hosting superconducting correlations would extend the scope of carbon-based topological phases into the superconducting regime.

Here we integrate triangulene Haldane chains on a  Au(111)/Nb(110) superconducting platform (Fig.~\ref{fig1}c) and probe the coupling of their fractional \spin12 edge states to the proximitized superconducting condensate at low temperatures. By comparing scanning tunnelling microscopy and spectroscopy measurements of chains with an increasing number of triangulene units, we identify a clear even-odd alternation in their low-energy spectral behaviour. Odd-length chains display YSR resonances originating from their fractional \spin12 edge states. Even-length chains, in contrast, exhibit only extra-gap singlet–triplet (ST) excitations reflecting antiferromagnetic coupling between the two edge states ~\cite{delgado2013,Mishra2021}.  
Theoretical modelling shows that this parity-dependent response originates from alternating magnetic ground states set by inter-edge exchange interactions, which determine the coupling of the edge spins to the superconducting host. These results demonstrate that the topological edge states of molecular Haldane chains remain robust in a superconducting environment.  More broadly, we establish proximitized gold films \cite{Wei2019b}  as a platform for engineering correlated spin–superconductor hybrids based on atomically precise carbon nanostructures~\cite{Huang2026}.

\textit{The superconducting Au(111) platform:}
We chose a Nb(110) single crystal as a bulk superconductor, and grew thin gold films on top using a layer of Ag atoms as a surfactant. Under ideal growth conditions \cite{VaxevaniPhD2026}, gold films exhibit the electronic and structural properties of Au(111) surfaces, namely, they appear with the characteristic herringbone reconstruction (Fig.~\ref{fig1}a) and Shockley surface state in scanning tunnelling microscopy and spectroscopy (STM/STS) measurements at low temperature (Supplementary Section S1). Owing to their thickness, smaller than the coherence length of the underlying superconductor, the films also host proximity-induced superconducting correlations. Low energy spectra such as the one shown in Fig.~\ref{fig1}d reproduce the bulk superconducting gap of the underlying Nb(110) substrate with a pair of sharp de Gennes-Saint James (dGSJ) bound states~\cite{deGennes1963} inside the zero conductance superconducting gap, characteristic of ballistic proximitized superconductors~\cite{Vaxevani2022extending,schneider_proximity_2023}. These dGSJ states serve as a spectroscopic marker of the proximity-induced superconducting gap~\cite{Ortuzar2023,Trivini2024}.

\begin{figure*}[t]
        \begin{center}
			\includegraphics[width=\textwidth]{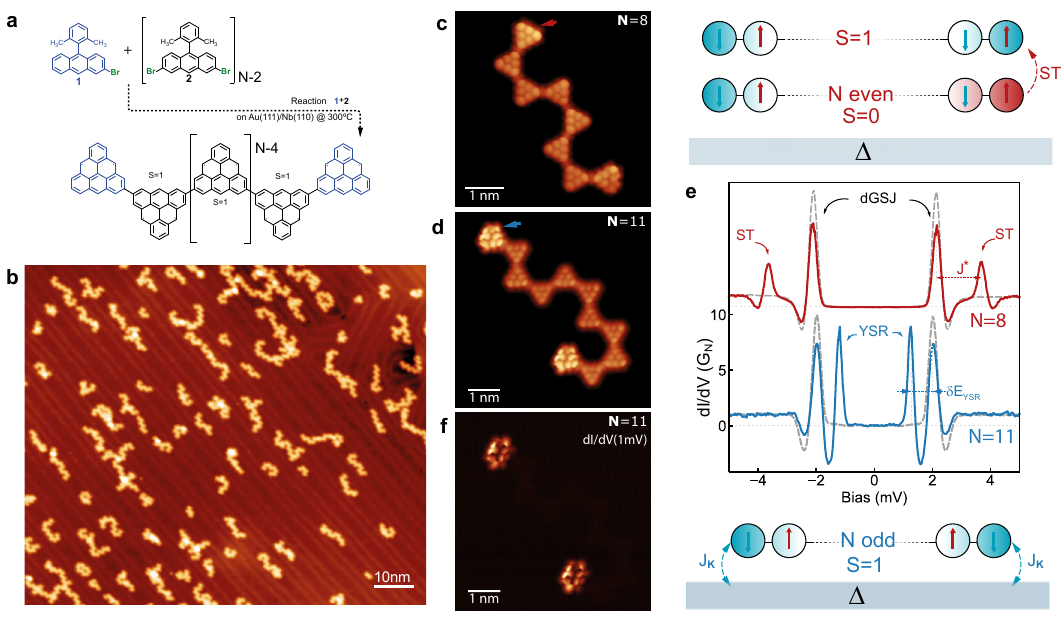}    
        \end{center}
			\caption{\textbf{Triangulene chains over a proximitized Au(111) superconductor}. 
            \textbf{a} Chemical structure of the organic precursors used \cite{Mishra2021} and the targeted triangulene chain.
            \textbf{b} STM image of the surface of a Au(111) film grown on Nb(110) substrates (Section S1 for details). 
            \textbf{c,d} Constant-height bond-resolved current images of two selected triangulene chains with N=8 and 11 units. These images and others in Supplementary Figures S3-S14 have been measured at low sample voltage (V=5 mV) using a CO-functionalized superconducting STM tip.   
            \textbf{e} Comparison of differential conductance spectra acquired at the termini of the two chains, using a superconducting STM tip to enhance the energy resolution at 1.2~K (\didv @ V = 5 mV, I = 500 pA). Even N chain shows two extra-gap singlet-triplet (ST) excitation peaks (scheme above); odd N chain shows sub-gap YSR resonances at energy J$^*_K$ below the dGSJ peaks, due to the interaction of end states with the superconductor (scheme below).
            \textbf{f} Constant-height \didv map of the N=11 chain at the sub-gap bias of its positive YSR peak. Only the ends f the N=11 triangulene chains show \didv signal, picturing the orbital character of the fractional end states. } 
		\label{fig2}
		\end{figure*}

\textit{Synthesis of triangulene chains on proximitized gold films:} 
The gold films also withstand the elevated temperatures required for on-surface synthesis of the triangulene chains \cite{Mishra2021}. Stepwise annealing of the molecular precursors [Fig.~\ref{fig2}a] on the Au(111)/Nb(110) substrate, induces their Ullmann coupling followed by cyclodehydrogenation to form triangulene chains of varying length across the Au(111) herringbone reconstruction (Fig.~\ref{fig2}b). Importantly, the reconstruction and the superconducting gap are unaltered by the annealing procedure~\cite{VaxevaniPhD2026}.
Constant-height current images of the chains using a CO-functionalized STM tip~\cite{gross2011high,Li2019}, as in   Figs.~\ref{fig2}c and \ref{fig2}d, reveal intact chains of $N$ covalently linked triangulene units, demonstrating the successful bottom-up engineering of graphene architectures directly on a proximitized superconducting substrate. 

Despite their similar shape, the chains in panels~\ref{fig2}c and \ref{fig2}d exhibit distinct low-energy spectra  at their termini. Figure~\ref{fig2}e compares differential conductance spectra (\didv) acquired on both chains (arrows in \ref{fig2}c,d). The odd-length chain ($N=11$) displays pronounced resonances inside the superconducting gap (Fig. S11), identified as YSR bound states induced by the coupling of localized edge spins to the proximitized superconducting surface~\cite{Ortuzar2023}. 
Spatially resolved maps of the YSR spectral weight (Fig.~\ref{fig2}f) show that these states are confined to the terminal triangulene units with a  characteristic shape that resembles radical states at zigzag edges \cite{Li2021}. This observation is consistent with spin-$\tfrac{1}{2}$  edge states at the ends of chains with open boundaries predicted by a valence-bond-solid description of an $S=1$ Haldane chain~\cite{affleck1987}, thus establishing the topological origin of the observed YSR states \cite{Tasaki2025,Aligia2025}.

In contrast, \didv spectra on the even-length ($N$=8) chain do not exhibit sub-gap resonances. Instead, two sharp peaks appear just above the superconducting gap, localized at the chain termini (Fig. S8). Spin-models of finite chains predict that  fractional spins may couple with each other, opening a ST gap that inversely scales with the chain length, an exponentially decaying correlation length (Section S6 and  \cite{delgado2013,Aligia2025}). Thus, we attribute these extra-gap features to inelastic edge-spin excitations, appearing as replicas of the dGSJ peaks shifted by the singlet-triplet transition energy \cite{Heinrich13a,Vaxevani2022extending}.

\begin{figure*}[t]
        \begin{center} 
			\includegraphics[width=\textwidth]{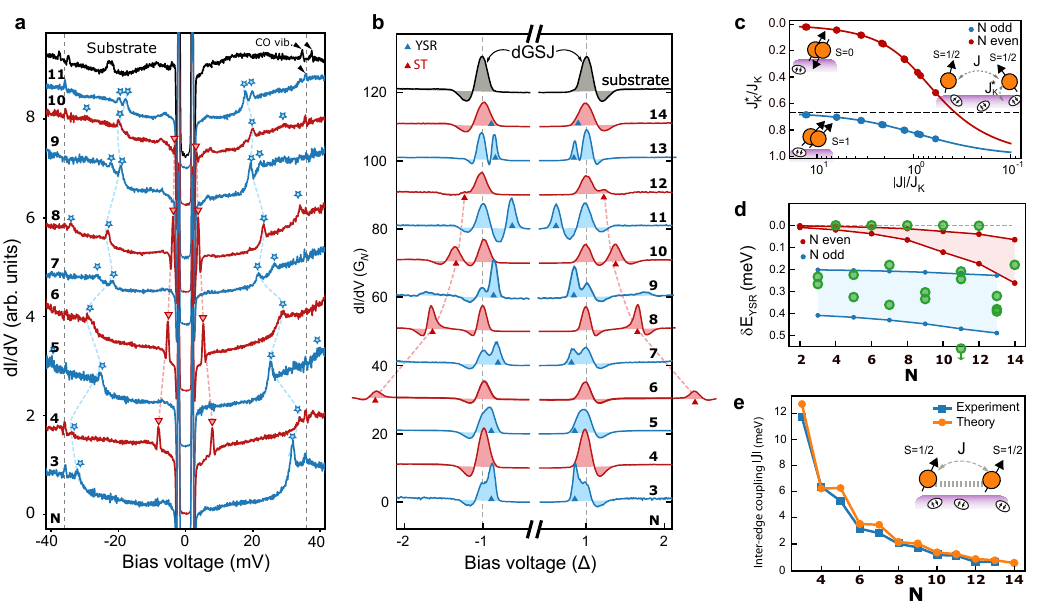}    
        \end{center}
			\caption{\textbf{Scaling of edge excitations with spin-chain length:}
            \textbf{a} Differential conductance spectra at the edge of triangulene chains with increasing length $N$. Down-pointing red triangles mark edge excitations; blue stars mark collective spin modes (\didv @ V = 50 mV, I = 1 nA).
            \textbf{b} Low-energy spectra at the termini of chains in panel a) and longer ones (Figs.~S3--S14). The dGSJ resonances are aligned to remove tip-dependent gap variations; YSR states are marked with blue up-pointing triangles (\didv @ V = 5 mV, I = 500 pA).
            \textbf{c} Parity-dependent renormalization of $J^*_K$ with $\lvert J\rvert$, from the two-site model of Section~S5. Even chains show negligible $J^*_K$ for larger $\lvert J\rvert$ values, consistent with the absence of YSR states in shorter chains.
            \textbf{d} YSR binding energy $\delta E_{\mathrm{YSR}}(N)=E_{\mathrm{dGSJ}}-E_{\mathrm{YSR}}(N)$ versus chain length $N$. For even chains, YSR states merge with dGSJ features, indicating negligible $J^*_K$; odd chains show YSR states for all lengths, with binding energy fluctuating due to variations $J_K$. A few cases were found outside the mapped range, e.g. Fig. S11. Symbols and shaded regions show model predictions using $J_K = 0.6 \pm 0.2$~meV, averaged from chains with a single edge state (Section~S4). Incipient YSR signal in even chains appears only for $N \geq 14$ (Fig.~S14).
            \textbf{e} Experimental and theoretical $\lvert J\rvert$ versus $N$. Theory: bilinear-biquadratic model of Section~S6. Experiment:  data from Figs.~S3--S14, including weak triplet-singlet excitations in odd chains.   }
             \label{fig3}
\end{figure*}

\textit{Length scaling of spin and YSR excitations:} 
The spectral differences between chains extend systematically to other lengths, alternating with the  parity of the chain length $N$, a signature of parity-dependent coupling between edge states.
Figure~\ref{fig3}a plots a set of wider-bias \didv spectra  acquired at the termini of triangulene chains ranging from $N$=3 to $N$=11. In this range of chain lengths, the plot reveals ST edge excitations over the ends of even-$N$ chains (red triangular markers in Fig.~\ref{fig3}a and \ref{fig3}b),  with  excitation energy that decreases systematically with chain length due to the decreasing overlap of end states through the bulk of the chain. For odd-$N$ chains, the corresponding spin excitation (triplet-singlet, TS) is much weaker (see Figs. S5-S13).

The complementary trend is observed for sub-gap \didv spectra across chain lengths from $N=3$ to $N=14$ (Fig.~\ref{fig3}b). Odd-$N$ chains exhibit YSR resonances at the termini (blue triangles), split off from the dGSJ features, whereas such states are absent in even-$N$ chains. The presence of YSR states in odd-length chains reveals a finite Kondo exchange coupling $J_K$ between edge states and the superconducting substrate, establishing their magnetic origin. Their binding energy, $E_{\mathrm{dGSJ}}-E_{\mathrm{YSR}}$, fluctuates within the gap due to site-dependent variations of $J_K$ across the substrate \cite{Franke11}, probably due to the Au(111) herringbone reconstruction \cite{Turco2024}. Even-$N$ chains display a featureless superconducting gap alongside extra-gap ST excitations, consistent with a singlet ground state decoupled from the superconducting condensate.

\textit{Two-site spin-superconductor model:} The parity-dependent coupling to the superconductor reflects distinct spin ground states for each parity of $N$, as anticipated in Fig.~\ref{fig1}a: parallel alignment of edge spins in odd-$N$ chains stabilizes a spin-triplet ground state with net magnetic moment, whereas antiparallel alignment in even-$N$ chains favours an open-shell singlet. Upon adsorption on the superconducting substrate, the exchange coupling $J_K$ becomes renormalized by the inter-edge interaction $J$ (Fig.~\ref{fig3}c), resulting in a parity-dependent effective coupling $J^*_K$ that accounts for the observed spectral alternation.

To model this behaviour, we describe edge states as two \spin12 impurities interacting with strength $J$, each coupled to a single-site superconductor ~\cite{Oppen2021,Ortuzar2023,trivini2023cooper,Huang2026}  with Kondo-like coupling $J_K$\footnote{For simplicity, in this explanation we assume both Kondo couplings to be equal and amounting to $J_K$. The real triangulene chains  can host distinct couplings on every end.}. In the limit of $|J|\gg J_K$, the spin chain behaves as a single $S=0$ (even $N$ chains) or $S=1$ (odd $N$ chains) impurity coupled to the superconductor with a renormalized Kondo coupling $J^*_K$ approaching $J^*_K=0$ and $J^*_K=2J_K/3$ for $S=0$ and $S=1$ ground states, respectively (Fig.~\ref{fig3}c and Section S5).  As the chain lengthens,  $|J|$ decreases and the system evolves toward two non-interacting spin-$\tfrac{1}{2}$ impurities, with $J^*_K$ approaching $J_K$. 

We validated this picture by comparing the parity-dependent evolution of the measured YSR binding energies, $\delta E_{\mathrm{YSR}}(N)=E_{\mathrm{dGSJ}}-E_{\mathrm{YSR}}(N)$  (green dots in Fig.~\ref{fig3}d), with predictions of the  two-site model for $J^*_K$ using $J_K$ and $J$ as parameters (Section S5). To estimate $J_K$ in the non-interacting limit, we selectively passivated one edge in several chains by tip-induced dehydrogenation \cite{Zhao2024} and measured spectra at the opposite end (Figs. S15  and S16). Removing one edge spin quenches inter-edge coupling $J$ (Fig. S16) and releases a spin at the opposite end, which systematically shows YSR states at lower energies, with average $J_K=0.6 \pm 0.2$~meV. The  inter-edge coupling  values $|J|$ were extracted from the ST and TS transitions plotted in Fig.~\ref{fig3}e.
With these parameters, the model reproduces the experimental evolution well: all odd-length chains host bound YSR states, distributed around the predicted binding energies, while even-length chains show no sub-gap YSR signal. In even-length chains, emergent YSR signal can be already detected for $N$=12 (Fig. S12) and $N$=14 (Fig. S14), making a  length regime at which $|J|\sim J_K$ and edge states enter the non-interacting regime.  

\begin{figure*}[t]
        \begin{center}
			\includegraphics[width=\textwidth]{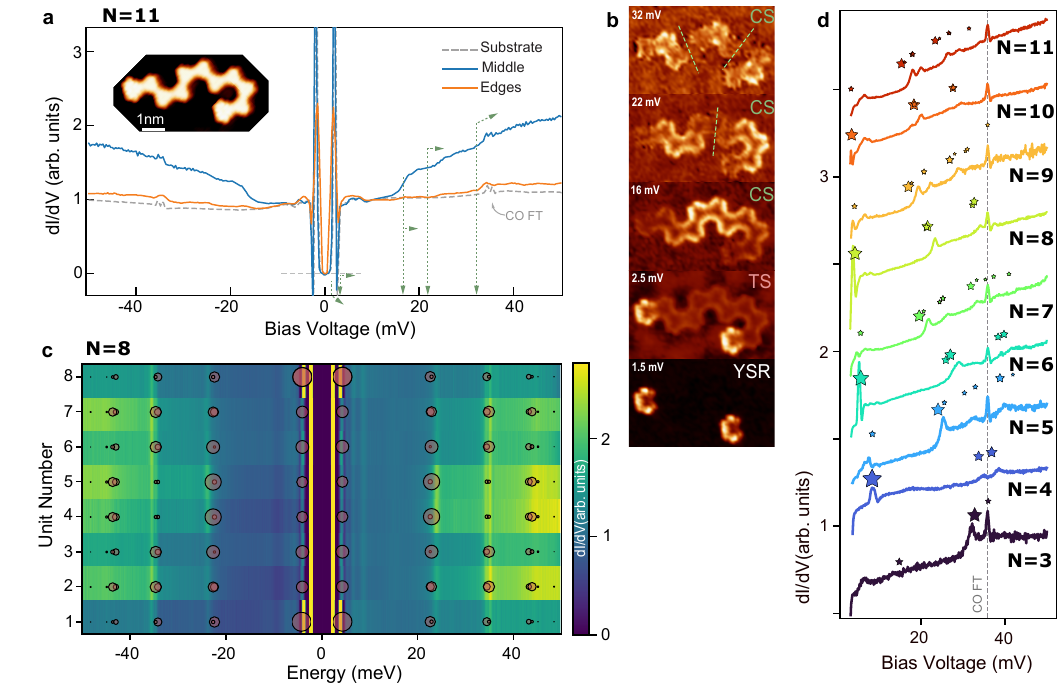}
        \end{center}
			\caption{\textbf{Collective spin modes in the bulk of triangulene chains.}
        \textbf{a} Long-range \didv spectrum on the terminal triangulene (orange) of the $N=11$ chain of Fig.~\ref{fig2}d, compared with a spectrum of inner triangulenes (blue). Both spectra are averaged over signal-rich regions in the zig-zag edges of end-triangulene and middle triangulenes, respectively. To facilitate comparison, spectra are normalized to match the same \didv value at V=10 mV. The dashed line is a reference spectrum averaged over the surface area. (\didv @ V = 50 mV, I = 1 nA).
        \textbf{b} \didv map obtained at the indicated voltage values, also marked in \textbf{a}. The maps are acquired from a grid of spectra at junction resistance 50 M$\Omega$.
        \textbf{c} \didv line profile obtained by stacking one spectrum on each triangulene unit of the $N=8$ chain in Fig.~\ref{fig2}c (each spectrum is the average of three on the three zig-zag edges of each triangulene). Spectra are normalized to their values in the 10–20 mV window. The circles overlaid on the line-profile are results from the BLBQ model of Section~S6: their area indicates the amplitude of each mode.
        \textbf{d} \didv spectra obtained by averaging $3 \times N$ spectra along triangulene chains for increasing $N$ (3 spectra per triangulene unit). The star size and position indicate amplitude and energy position of edge-spin and CS excitation modes calculated with the BLBQ model of Section~S6.                    }
		\label{fig4}
\end{figure*}

\textit{Excitation of collective spin modes:} 
The spectral sequence of Fig.~\ref{fig3}a also shows peaks at higher energy (blue stars) associated with collective spin (CS) modes of the chain, marking the onset of S=1 many-body spin excitations modes extending along the bulk~\cite{Spinelli2014,Drost2023,Zhao2024b}. These modes exhibit weaker amplitude at the chain edges and stronger signal in the bulk, appearing as increases in the conductance as shown in Fig.~\ref{fig4}a for the $N=11$ chain. 
Differential conductance maps at the energy of each mode (Fig.~\ref{fig4}b) reveal a characteristic pattern of standing waves with the number of nodes increasing with energy. The spectral map of Fig.~\ref{fig4}c displays the spatial distribution of edge-localized ST and bulk CS excitations along the $N=8$ chain of Fig.~\ref{fig2}c, also showing  \didv inelastic steps and a pattern of nodes as a function of excitation energy. 
The spatial distribution of both ST and CS excitations is perfectly reproduced by a Heisenberg Hamiltonian within the bilinear-biquadratic (BLBQ) approximation~\cite{affleck1987,Mishra2021} (Section~S6). The observed patterns of mode amplitude (circles in Fig.~\ref{fig4}c) match those of confined many-body spin excitations of a free chain, decoupled from the superconductor by the larger Haldane gap.

To trace the evolution of these excitations with chain length and parity, we plotted in Fig.~\ref{fig4}d the averaged spectra for chains from $N=3$ to $N=11$. CS features alternate in number and position with length, in a parity-dependent pattern captured by theoretical models \cite{Machens2013,Mishra2021}. The model reproduces the energy lowering of ST/TS excitations with length (as in Fig.~\ref{fig3}e), due to decreasing wavefunction overlap, and the even-odd alternation of their intensity,  attributed to variations in ground-state spin multiplicity~\cite{delgado2013,Mishra2021}. The onset of CS excitations decreases exponentially following a Landé-like oscillating pattern~\cite{Alfredlande1923, Machens2013},  converging into the bulk Haldane gap of massive magnon excitations~\cite{Haldane1983}.

Our proximitized platform establishes a new class of hybrid system, where fractional $S=1/2$ boundary spins remain robust at the ends of $S=1$ triangulene chains and acquire a parity-dependent Yu-Shiba-Rusinov fingerprint of their topological fractionalization. The shown length-controlled switching of the magnetic ground state, read out through the presence or absence of sub-gap YSR excitations, constitutes a new mechanism for encoding and detecting spin correlations at the molecular scale.  We foresee that extending this approach to other designer spin systems in carbon-based $\pi$-conjugated architectures could give access to regimes where entanglement between fractional edge spins and screening by the superconducting condensate coexist and compete~\cite{Huang2026}, opening a route toward topologically protected molecular spin qubits and novel hybrid quantum devices based on carbon nanostructures.

 \nolinenumbers

\begin{acknowledgments}
This research was funded by  MCIN/AEI/10.13039/501100011033 and the European Regional Development Fund (ERDF) through Grants No. PID2022-140845OB-C61, PID2022-140845OB-C62, and CEX2020-001038-M,  and by 
the European Union Horizon 2020 and HE programs through FET-Open project SPRING (No. 863098), the ERC Synergy Grant MolDAM (No. 951519), and the ERC-AdG CONSPIRA (No. 101097693). 
K.V. acknowledges support from MCIN/AEI/10.13039/501100011033 through PhD scholarship PRE2022-102887.
J.O. acknowledges support from the Basque Government through PhD scholarship No. PRE\_2021\_1\_0350.
D.W. acknowledges support from "Juan de la Cierva" grant FJC2020-043831-I funded by MCIN/AEI/10.13039/501100011033 and by the European Union NextGenerationEU/PRTR, as well as from NSFC project No. 12474474.

\end{acknowledgments}

\bibliographystyle{apsrev4-1} 

%

\end{document}


\title{Supplementary Information:  \\\rev{Parity-dependent coupling of  molecular spin chains \\to a superconductor}   }

\author{Katerina Vaxevani}\thanks{These authors contributed equally to this work}
    \affiliation{CIC nanoGUNE-BRTA, 20018 Donostia-San Sebasti\'an, Spain}

\author{Jon Ortuzar}\thanks{These authors contributed equally to this work}
      \affiliation{CIC nanoGUNE-BRTA, 20018 Donostia-San Sebasti\'an, Spain}
  \affiliation{Quantronics group, Service de Physique de l'\'Etat Condens\'e \mbox{(CNRS, UMR 3680)}, IRAMIS, CEA-Saclay, Universit\'e Paris-Saclay, 91191 Gif-sur-Yvette, France}
  
\author{Stefano Trivini}\thanks{These authors contributed equally to this work}
    \affiliation{CIC nanoGUNE-BRTA, 20018 Donostia-San Sebasti\'an, Spain}

\author{Georg Monninger}
    \affiliation{CIC nanoGUNE-BRTA, 20018 Donostia-San Sebasti\'an, Spain}

\author{Dongfei Wang}
    \affiliation{CIC nanoGUNE-BRTA, 20018 Donostia-San Sebasti\'an, Spain}
    \affiliation{School of Physics, Beijing Institute of Technology, Beijing 100081, China}
 
\author{Manuel Vilas-Varela}
    \affiliation{Centro Singular de Investigaci\'on en Qu\'imica Biol\'oxica e Materiais Moleculares (CiQUS) and Departamento de Qu\'imica Org\'anica, Universidad de Santiago de Compostela, 15782 Santiago de Compostela, Spain}
    
\author{Lucía Gómez-Rodrigo}
    \affiliation{Centro Singular de Investigaci\'on en Qu\'imica Biol\'oxica e Materiais Moleculares (CiQUS) and Departamento de Qu\'imica Org\'anica, Universidad de Santiago de Compostela, 15782 Santiago de Compostela, Spain}

\author{Diego Peña}
    \affiliation{Centro Singular de Investigaci\'on en Qu\'imica Biol\'oxica e Materiais Moleculares (CiQUS) and Departamento de Qu\'imica Org\'anica, Universidad de Santiago de Compostela, 15782 Santiago de Compostela, Spain}
    \affiliation{Oportunius, Galician Innovation Agency (GAIN), 15702 Santiago de Compostela, Spain}  

\author{Jose Ignacio Pascual} \email{ji.pascual@nanogune.eu}
    \affiliation{CIC nanoGUNE-BRTA, 20018 Donostia-San Sebasti\'an, Spain}
    \affiliation{Ikerbasque, Basque Foundation for Science, 48013 Bilbao, Spain}
\setstretch{1.1} 
\date{\today}
\maketitle
\tableofcontents

\setstretch{1.0} 

\onecolumngrid\newpage
 
\subsection{Experimental  Methods}

The experiments were performed in a Joule-Thomson low-temperature scanning tunnelling microscope (Specs GmbH) operating at a base temperature of 1.2 K under ultra-high vacuum conditions. To induce superconducting correlations in gold while preserving the chemical reactivity, surface density of states and structure of bulk Au(111), we grew thin gold films on a bulk Nb superconductor. Nb(110) single crystals were cleaned by repeated cycles of Ar$^+$ sputtering and annealing until extended flat terraces were obtained \cite{Wang_epitaxial_2025}, and Au layers were subsequently grown on the room-temperature Nb(110) surface precovered with an Ag layer acting as a surfactant \cite{VaxevaniPhD2026}.

The experiments reported here were performed on Au films of about 25--30 monolayers, exhibiting extended flat terraces with the characteristic herringbone reconstruction of Au(111) and a well-defined Shockley surface-state onset at $\approx -500$ meV (Fig.~S1). These observations indicate that both the surface structure and the intrinsic electronic properties are those of a bulk Au(111) surface.

To grow the triangulene chains, the molecular precursors shown in Fig.~2a were deposited in a 3:1 ratio  on the gold films held at room temperature. The sample was subsequently annealed stepwise, first to 200~$^{\circ}$C to induce dehalogenation and Ullmann C--C coupling of the precursors, and then to 340~$^{\circ}$C to induce their cyclodehydrogenation into the target triangulene chains.

Differential conductance (\didv) spectra were acquired using Nb superconducting tips to enhance the energy resolution \cite{Franke11}. Nb tips were prepared by repeated strong indentations of a clean W tip into a clean Nb(110) crystal. In most measurements, bond-resolved imaging at constant height and \didv spectroscopy were performed with CO-functionalized tips. All \didv spectra were measured with a lock-in amplifier, using an AC modulation of the sample bias of either 50 $\mu$V or a few mV's, depending on the voltage window studied.

\begin{figure*}[b]
        \begin{center}
			\includegraphics[width=0.8\textwidth]{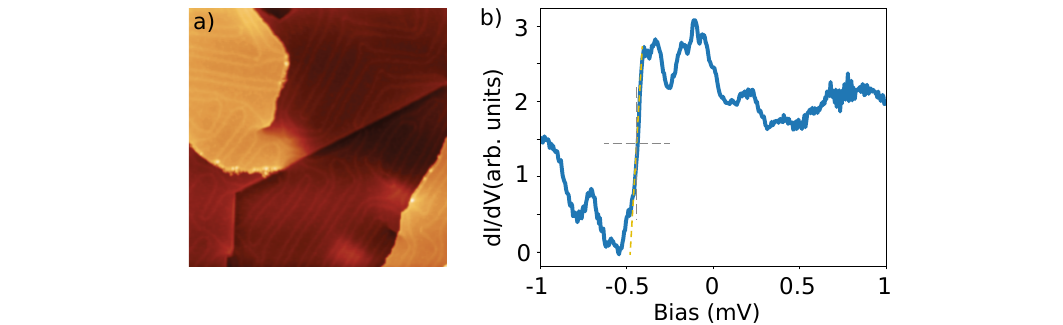}    
        \end{center}
			\caption{  (Left) STM image of the surface of a 25 ML Au(111) film, exhibiting the herringbone reconstruction and (right) \didv spectra measured at 1.2 K, showing the characteristic onset of its surface state close to -0.5 V sample bias. This value is close to the surface state onset on bulk Au(111) substrates. 
   }
		\label{SFig-N8-LS}
		\end{figure*}

\clearpage

\subsection{Deconvolution of the superconducting tip gap and YSR bound state energies}

Tunnelling in an asymmetric SIS junction can produce negative differential resistance (NDR) \cite{Devyatov1998,Lyo1989,Franke2011b}, shifting the apparent energy of in-gap resonances from their intrinsic position. Deconvolution of the tip DOS is therefore necessary, but on a proximitized substrate the superconducting DOS is not well described by a Dynes function \cite{Pillet2010a}. We instead modelled both tip and sample using Arnold-type spectral functions \cite{arnold1978theory} (the tip frequently is covered by Au after small indentations for \textit{in situ} sharpening and, thus, is considered usually as a proximitized electrode), fitting first a reference SIS spectrum on the bare surface to extract the tip DOS, and then the YSR spectra with the tip DOS fixed, recovering the intrinsic bound-state energies.

The Arnold density of states \cite{arnold1978theory} reads:

\begin{equation}
N_{\mathrm{Arnold}}(E)
=
-\mathrm{Im}\!\left[
\frac{
\displaystyle \frac{E}{\Omega_N(E)}
\left(
i\,F(E)\cos\phi(E)+\sin\phi(E)
\right)
}{
i\,F(E)\sin\phi(E)-\cos\phi(E)
}
\right]_{E\rightarrow E+i\eta},
\end{equation}

where

\begin{equation}
\Omega_N(E)=\sqrt{E^2-\Delta_N^2},
\qquad
\Omega_S(E)=\sqrt{E^2-\Delta_S^2},
\end{equation}

and

\begin{equation}
F(E)=\frac{E^2-\Delta_S\Delta_N}{\Omega_S(E)\,\Omega_N(E)},
\qquad
\phi(E)=R\,E.
\end{equation}

Here, \(R\) is a phase-like parameter that controls the detailed line shape of the coherence peaks (dGSJ quasiparticle peaks in our case), and \(\eta\) is a phenomenological intrinsic broadening introduced through the substitution \(E\rightarrow E+i\eta\). In the present implementation, the normal-side gap was set to zero, \(\Delta_N=0\), so that the model is fully determined by the three parameters \(\Delta_S\), \(R\), and \(\eta\).

\begin{figure*}[h]
        \begin{center}
			\includegraphics[width=\textwidth]{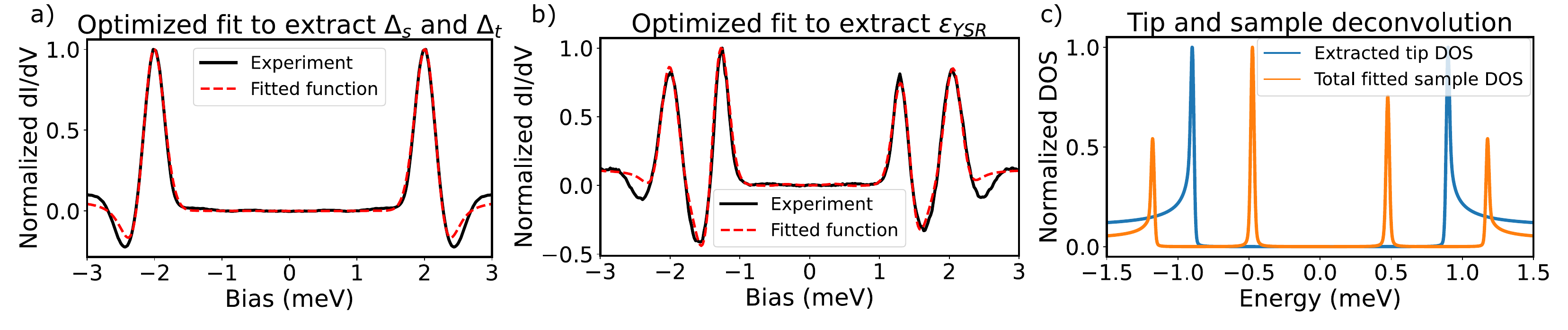}    
        \end{center}
			\caption{\textbf{Numerical extraction of the intrinsic YSR energies for N=11}(a) Normalized SIS reference spectrum (black) together with the optimized Arnold-based SIS fit (red dashed), used to extract the effective superconducting density of states of the tip.(b) Normalized YSR spectrum acquired with the same superconducting tip (black), fitted using the previously extracted tip DOS and a model sample DOS consisting of a superconducting continuum plus in-gap YSR states (red dashed). (c) Deconvoluted density of states used in the analysis, showing the extracted tip DOS (blue) and the fitted total sample DOS (orange). This procedure allows the intrinsic energetic position of the YSR states to be determined from the sample DOS rather than directly from the YSR peak in the measured SIS conductance.}
		\label{SFig2}
\end{figure*}

As shown in SFig.~\ref{SFig2}a (red dashed line), the convolution of two Arnold DOS functions reproduce the main SIS peak-dip structure of a experimental \didv spectrum on the bare Au(111)/Nb substrate (black line) and provides an effective intrinsic DOS for the superconducting tip (blue spectra in SFig.~\ref{SFig2}c) using a known $\Delta_{\mathrm{sample}}$.  The fit procedure to the spectra of a bare SIS (SFig. 2a) yields tip gap values of $\Delta_{\mathrm{tip}}=0.9$ meV using a precalibrated $\Delta_{\mathrm{sample}}=1.2$meV. The blue plot in Fig.~\ref{SFig2}c shows the corresponding tip's Arnold DOS. 

The fit captures perfectly the shape of the de Gennes-Saint James states (dGSJ)  \cite{deGennes1963} frequently appearing in tunnelling spectra. as sharp peaks accompanied by nearby dip-like features close to the SIS gap edge.  The dGSJ states are Andreev-like sub-gap states located close to the superconducting gap edge that can be interpreted as superconducting quasiparticle excitation features of the proximitized film \cite{Vaxevani2022extending, Ortuzar2023}. Convolution of their intricate shape results in the additional NDR, slightly shifting their position. 

The extracted tip DOS is then used in the analysis of spectra containing Yu-Shiba-Rusinov (YSR) resonances.   
To extract YSR bound state energy from spectra with YSR peaks with NDR, we merged the sample's Arnold-type DOS with Lorentzian peaks (with additional Gaussian broadening) at position $\epsilon_{YSR}$, representing the YSR resonances, and convoluted with the tip DOS within the standard SIS tunnelling formalism. The values of $\epsilon_{YSR}$ can then be obtained from a fit to the measured \didv spectrum,  shown in SFig.~\ref{SFig2}b. Fitting the YSR spectrum of the N=11 triangulene chain in Fig. 2 of the manuscript (Fig. S2b) we obtain  $\epsilon_{YSR}= 0.47$ meV, corresponding to a bound state energy of  $\delta^{sim} E_{YSR} =  \Delta_{sample}-\epsilon_{YSR}$=0.78 meV, independent of the   $\Delta_{sample}$ value used. 

The \textit{bound state energy} used in the manuscript, $\delta E_{YSR} = |E_{dGSJ}-E_{YSR}|$, is obtained directly from the voltage  difference of \didv peak positions, yielding $\delta^{exp} E_{YSR} =$0.8 meV, in close agreement with the value obtained from the fit, $\delta^{sim} E_{YSR} =$0.78 meV. This agreement comes because NDR-induced shifts affect both the dGSJ and YSR peaks equally and thus reduces the  effect in their difference, keeping the error below $\sim 3\%$. Therefore, the plots in Fig. 3d and 3e, uses peak values for $\delta^{exp} E_{YSR}$ and  $E_{ST}, E_{TS}$.  As shown in   section \ref{SI:expres}, many YSR states overlap with the dGSJ peaks; peak positions are then obtained by fitting two Lorentzian peaks to the \didv spectra.  

\clearpage

\subsection{Supplementary experimental results}  \label{SI:expres}

\begin{figure*}[h]
        \begin{center}
			\includegraphics[width=\textwidth]{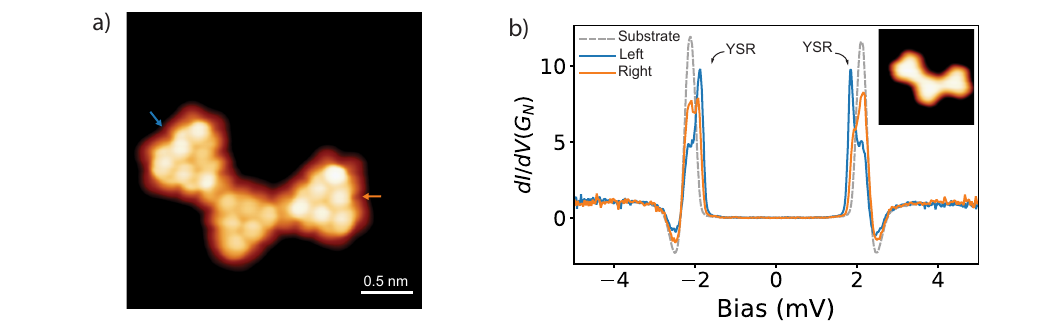}    
        \end{center}
			\caption{\textbf{Results for a N=3 [3]-triangulene chain:} a) Bond-resolved current image of triangulene chain with 3 units (V=5 mV, constant-height @ tunnel resistance R$_T \approx$16 M$\Omega$). b) Low energy \didv spectra    over the   termini. Dashed gray line is  a reference spectrum on the Au(111) substrate. Spectra on the chain termini show YSR  peaks developing inside the dGSJ excitations. }
		\label{SFig3}
\end{figure*}

\begin{figure*}[h]
        \begin{center}
			\includegraphics[width=\textwidth]{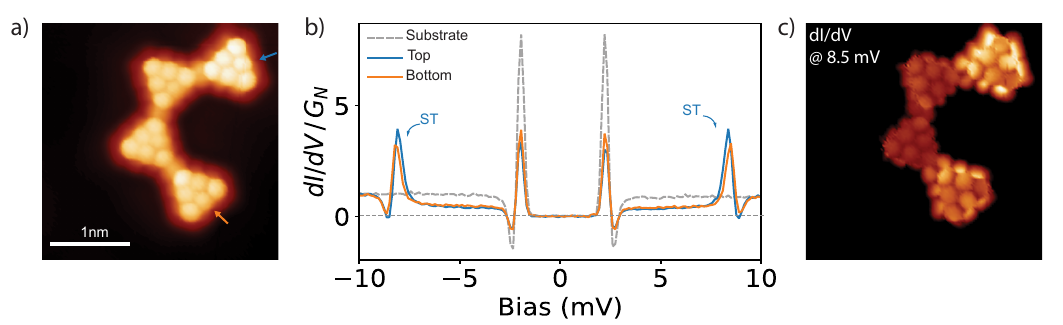}    
        \end{center}
			\caption{\textbf{Results for a N=4 [3]-triangulene chain:} a) Bond-resolved current image of triangulene chain with 4 units (V=8.5 mV, constant-height @ R$_T \approx$16 M$\Omega$). b) Low energy \didv spectra    over the   termini. Dashed gray line is  a reference spectrum on the Au(111) substrate. Spectra on the chain termini show peaks at $\sim$ 8.5 meV, corresponding to ST excitations with energy E$_{ST}$=6.3 mV. c) Differential conductance map at the peak energy, measured simultaneously to image in a), showing larger conductance over the edge triangulene units.  }
		\label{SFig4}
\end{figure*}

\begin{figure*}[h]
        \begin{center}
			\includegraphics[width=\textwidth]{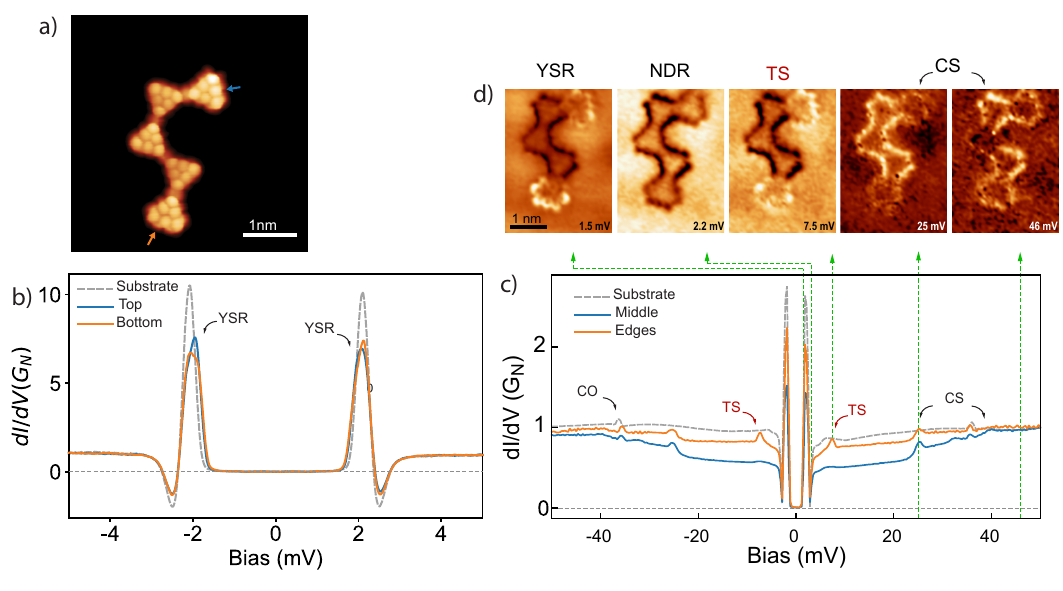}    
        \end{center}
			\caption{\textbf{Results for a N=5 [3]-triangulene chain:} a) Bond-resolved current image of triangulene chain with 5 units (V=5 mV, constant-height @ R$_T \approx$16 M$\Omega$). b) Low energy \didv spectra    over the   termini. Dashed gray line is a reference spectrum on the Au(111) substrate. Spectra on the chain termini show YSR states developing from the dGSJ peaks and TS excitation peaks, shown in c). c) Long range \didv spectra averaged at several position over the edge and the bulk of  the triangulene chain in a). d) Differential conductance maps of the triangulene chain in a). The maps are chosen at  bias values of the YSR peak, the TS excitation, and two collective spin excitation steps, respectively. Both c) and d) were obtained from a spectral grid over the whole triangulene chain.  }
		\label{SFig5}
\end{figure*}

\begin{figure*}[h]
        \begin{center}
			\includegraphics[width=\textwidth]{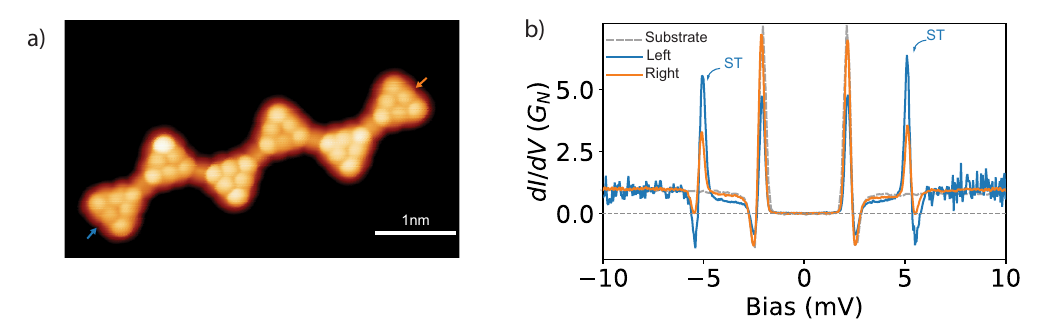}    
        \end{center}
			\caption{\textbf{Results for a N=6 [3]-triangulene chain:} a) Bond-resolved current image of triangulene chain with 6 units (V=5 mV, constant-height @ R$_T \approx$16 M$\Omega$). b) Low energy \didv spectra    over the   termini. Dashed gray line is  a reference spectrum on the Au(111) substrate. Spectra on the chain termini show ST excitation peaks with E$_{ST}$=3.1 mV. }
		\label{SFig6}
		\end{figure*}


        \begin{figure*}[h]
        \begin{center}
			\includegraphics[width=\textwidth]{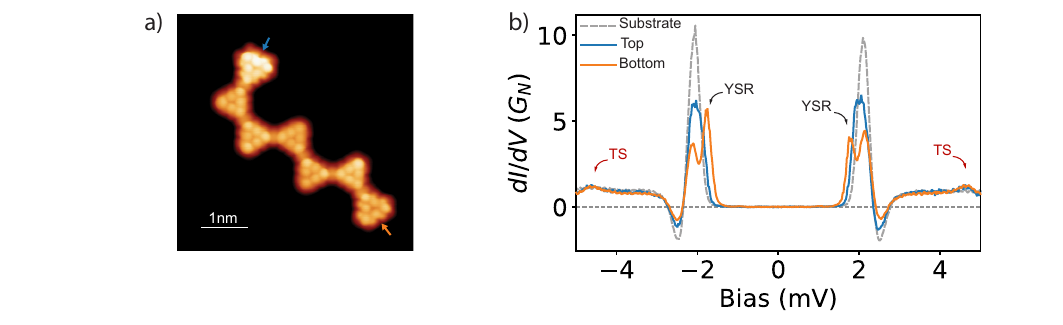}    
        \end{center}
			\caption{\textbf{Results for a N=7 [3]-triangulene chain:} a) Bond-resolved current image of triangulene chain with 7 units (V=5 mV, constant-height @ R$_T \approx$16 M$\Omega$). b) Low energy \didv spectra    over the   termini. Dashed gray line is  a reference spectrum on the Au(111) substrate. Spectra on the chain termini show YSR excitation peaks developing inside the gap between dGSJ peaks. Weak TS excitation peaks are observed outside the gap, with E$_{TS}$=2.7 mV.}
		\label{SFig7}
\end{figure*}

\begin{figure*}[h]
        \begin{center}
			\includegraphics[width=\textwidth]{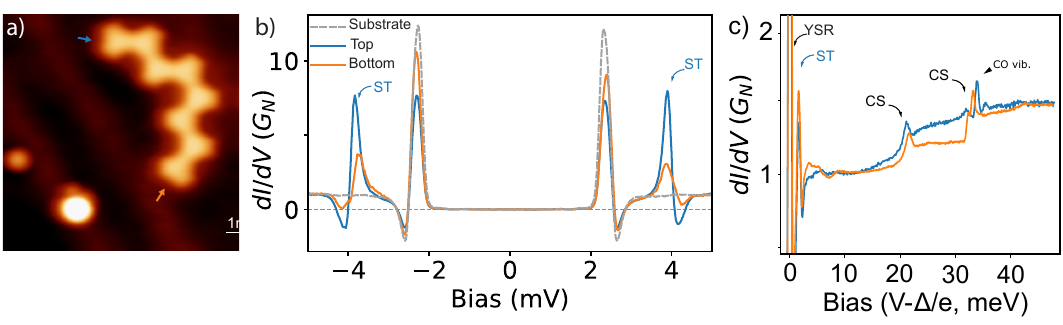}    
        \end{center}
			\caption{\textbf{Results for a N=8 [3]-triangulene chain:} a) Constant current image of triangulene chain with 8 units (V=5 mV). b) Low energy \didv spectra  over the   termini. Dashed gray line is a reference spectrum on the Au(111) substrate. Spectra on the chain termini show ST excitation peaks with E$_{ST}$=1.9 mV. \textcolor{black}{c) Comparison of  spin excitation spectra for two distinct N=8 chains. Both host the many-body S=1 spin excitations in the same frequencies. The peak at $\sim$35 meV is related to the  vibrations of the CO at the STM tip.} }
		\label{SFig8}
		\end{figure*}


        \begin{figure*}[h]
        \begin{center}
			\includegraphics[width=\textwidth]{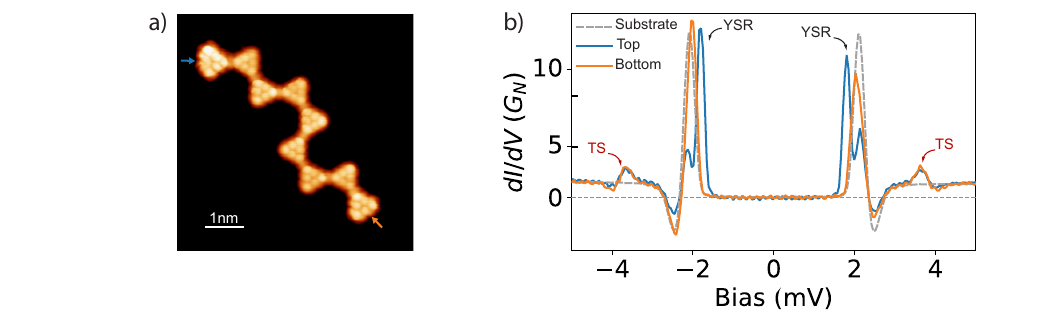}    
        \end{center}
			\caption{\textbf{Results for a N=9 [3]-triangulene chain:} a) Bond-resolved current image of triangulene chain with 9 units (V=5 mV, constant-height @ R$_T \approx$16 M$\Omega$). b) Low energy \didv spectra over the   termini. Dashed gray line is  a reference spectrum on the Au(111) substrate. Spectra on the chain termini show YSR excitation peaks developing inside the dGSJ excitations. Weak TS excitation peaks are observed outside the gap, with with E$_{TS}$=1.6 mV.}
		\label{SFig9}
\end{figure*}
  

\begin{figure*}[h]
        \begin{center}
			\includegraphics[width=\textwidth]{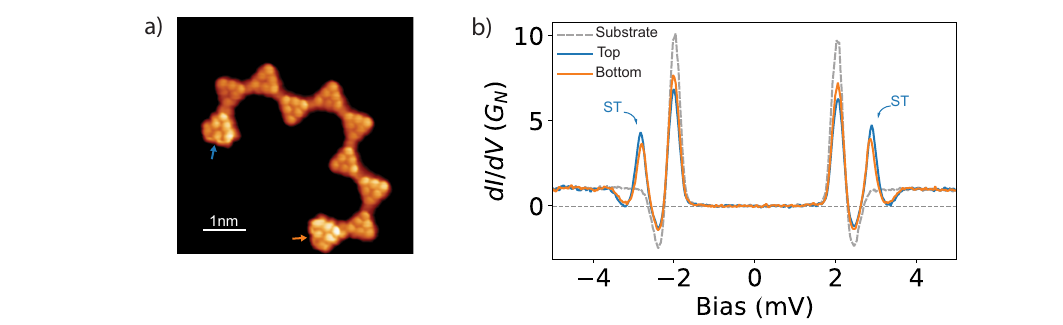}    
        \end{center}
			\caption{\textbf{Results for a N=10 [3]-triangulene chain:} a) Bond-resolved current image of triangulene chain with 10 units (V=5 mV, constant-height @ R$_T \approx$16 M$\Omega$). b) Low energy \didv spectra    over the   termini. Dashed gray line is  a reference spectrum on the Au(111) substrate. Spectra on the chain termini show ST excitation peaks with E$_{ST}$=1.0 mV. }
		\label{SFig10}
\end{figure*}


\begin{figure*}[h]
        \begin{center}
			\includegraphics[width=\textwidth]{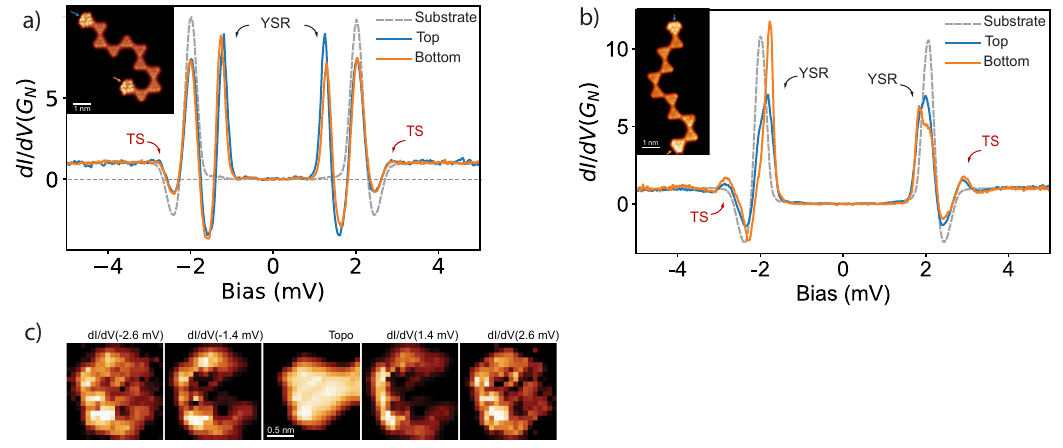}    
        \end{center}
			\caption{\textbf{Results for a N=11 [3]-triangulene chain:} a) Low energy \didv spectra  over both termini of the N=11 triangulene chain in Fig. 2d. b)    Low energy \didv spectra  over both termini of another N=11 triangulene chain with less YSR bound energy $\delta E_{YSR}$. In both plots: inset are bond resolved images and dashed gray lines reference spectra on the Au(111) substrate.  Both chains show weak TS excitation shoulders as marked, with E$_{TS}\sim$0.8 mV, the one in b) with larger peak amplitude. \rev{c) Spatial distribution of YSR peak amplitude and TS onset over the top-end triangulene in a), obtained from 20$\times$20 spectral grid. Both signals are distributed around the zigzag periphery, with similar shape at both polarities. } }
		\label{SFig11}
\end{figure*}


\begin{figure*}[h]
        \begin{center}
			\includegraphics[width=\textwidth]{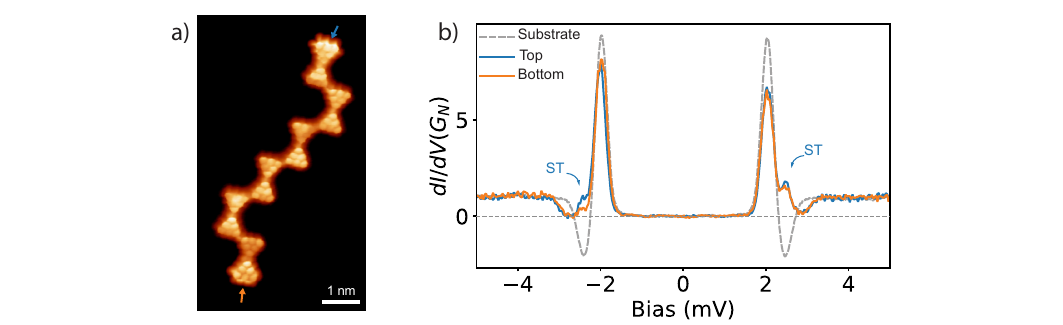}    
        \end{center}
			\caption{\textbf{Results for a N=12 [3]-triangulene chain:} a) Bond-resolved current image of triangulene chain with 12 units (V=5 mV, constant-height @ R$_T \approx$16 M$\Omega$). b) Low energy \didv spectra    over the   termini. Dashed gray line is  a reference spectrum on the Au(111) substrate. Spectra on the chain termini show faint ST excitation peaks with E$_{ST}\sim$0.5 mV. \rev{The dGSJ resonances appear with particle-hole asymmetry characteristic of YSR states with finite Coulomb potential scattering. Here, we attribute it to the incipient emergence of YSR states, still overlapping with the film's quasiparticle peaks.  } }
		\label{SFig12}
\end{figure*}


\begin{figure*}[h]
        \begin{center}
			\includegraphics[width=\textwidth]{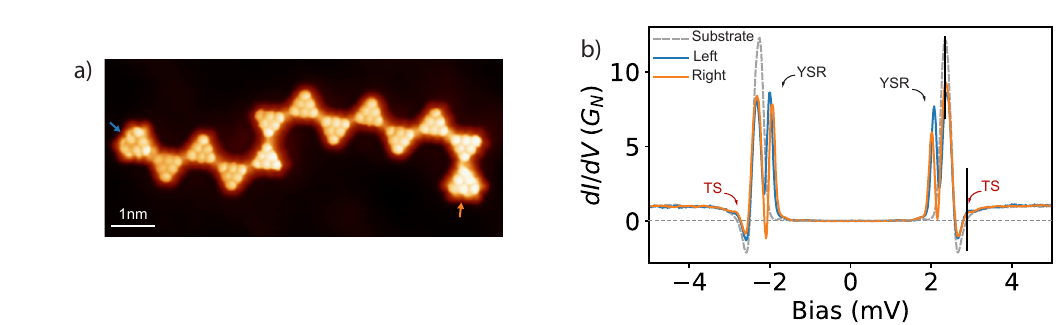}    
        \end{center}
			\caption{\textbf{Results for a N=13 [3]-triangulene chain:} a) Bond-resolved current image of triangulene chain with 13 units (V=5 mV, constant-height @ R$_T \approx$16 M$\Omega$). b) Low energy \didv spectra    over the   termini. Dashed gray line is  a reference spectrum on the Au(111) substrate. Spectra on the chain termini show YSR excitation peaks developing inside the dGSJ excitations. Spectra on the chain termini show faint TS excitation peaks with E$_{TS}\sim$0.5 mV.}
		\label{SFig13}
\end{figure*}


\begin{figure*}[h]
        \begin{center}
			\includegraphics[width=\textwidth]{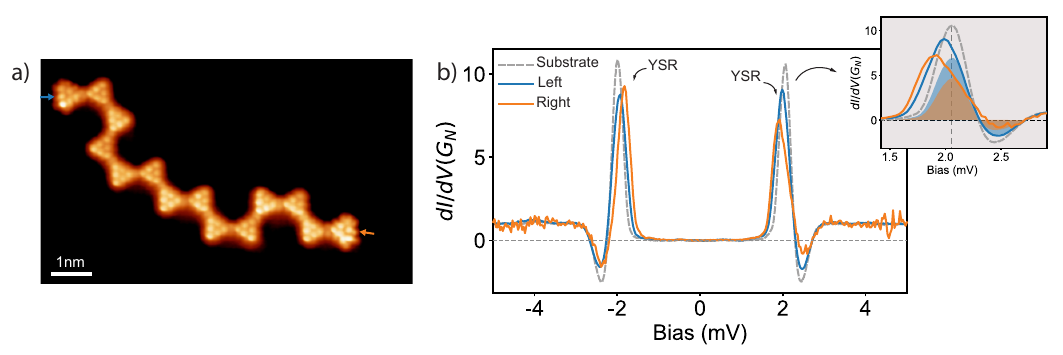}    
        \end{center}
			\caption{\textbf{Results for a N=14 [3]-triangulene chain:} a) Bond-resolved current image of triangulene chain with 14 units (V=5 mV, constant-height  @ R$_T \approx$16 M$\Omega$). b) Low energy \didv spectra    over the   termini. Dashed gray line is a reference spectrum on the Au(111) substrate. In the inset, we show the scaled components of the dGSJ states for each termini, to emphasize the peak components due to YSR state.}
		\label{SFig14}
\end{figure*}

 \clearpage

\newpage

\textbf{Tip-induced removal of fractional edge states:}  To remove intramolecular coupling between edge states,  we selectively dehydrogenate one site of a terminal triangulene by applying controlled voltage pulses over a preselected site, as shown in ref. \cite{Zhao2024}. The removal of one hydrogen atom exposed a $\sigma$-radical of the C site, which likely binds to the Au surface underneath. In these cases, the modified terminal triangulene  appears in STM images with a characteristic orbital shape shown in the inset of Fig. S15a. This characteristic shape is taken as a fingerprint of the target modification of the end triangulene.  

\begin{figure*}[h]
        \begin{center}
			\includegraphics[width=\textwidth]{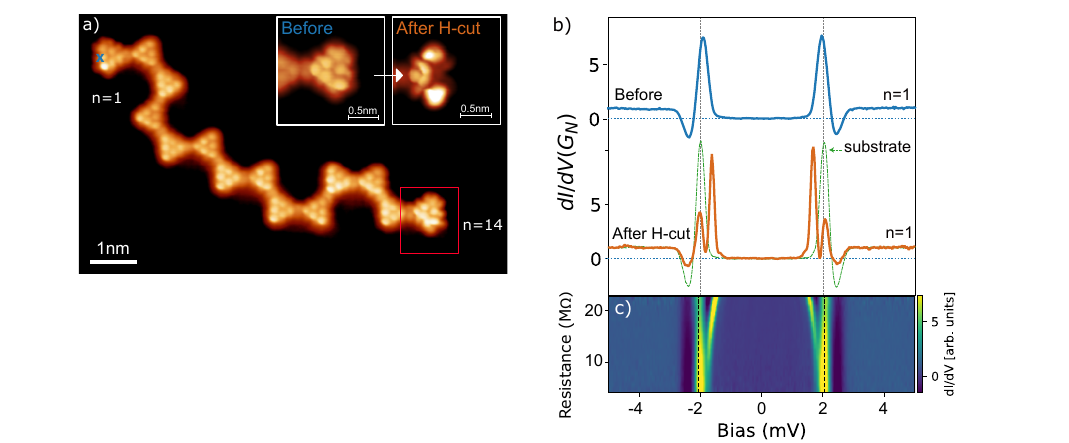}    
        \end{center}
			\caption{\textbf{Edge state passivation by dehydrogenation of a terminal triangulene - N=14. }
             a) Bond-resolved current STM image of the $N$=14 chain. The insets show a zoom image of the  \textbf{N}=14 termini, before and after applying a bias pulse over the marked C site. The pulse produces dehydrogenation of the targeted C edge site, after which the triangulene appears with  characteristic orbital shape in the images ($V$=5 mV).
            b) Differential conductance spectra on the \textbf{N}=1 termini before and after the dehydrogenation in a). The YSR peak of Fig. S14 over the \textbf{N}=1 triangulene shifts to a binding energy $\delta E_{\mathrm{YSR}}=0.36$ meV  when the opposite end is dehydrogenated (V = 5 mV, I = 500 pA).
            c) Differential conductance spectral map vs tip-sample resistance over the left triangulene. The YSR state shifts towards the gap edges,  and merges with dGSJ resonances,  as the tip approaches, indicating that the edge spin lies in the weak interaction regime ($J_K<<\Delta$).}
		\label{deH}
		\end{figure*}

The spectral signature of a YSR state alone does not directly reveal the parity of the underlying ground state, since sub-gap peaks may correspond either to a quasiparticle excitation of a BCS ground state or to a BCS-like excitation of a bound-quasiparticle ground state \cite{Heinrich2018}. These two regimes, defined by weak ($J_K<<\Delta$) and strong ($J_K>>\Delta$) exchange coupling respectively, are separated by a parity-changing quantum phase transition. To determine the interaction regime, we measured the evolution of the YSR resonance as the STM tip approached to the edge state of the dehydrogenated \textit{N}=14 chain of Fig.~\ref{deH}. As shown previously~\cite{Heinrich15,trivini2023cooper}, at the initial stages of the approach to the molecule, the tip exerts attractive forces that weaken the molecule–substrate interaction, e.g. reduce $J_K$.  On the ends of the triangulene chains, this perturbation shifts the YSR energy toward the superconducting gap edge in the weak-coupling limit, or towards zero-energy, the quantum phase transition point, in the strong-coupling regime. For a CO-functionalized STM tip, attractive forces become largest for junction resistances in the range of 10 M$\Omega$ to 20~M$\Omega$ \cite{Corso2015}, and become repulsive at  larger conductance. In Fig.~\ref{deH}c, we show the evolution with tip approach of the YSR states on the $N$=1 end of the triangulene shown in Fig.~S15a. The observed shift of YSR states toward the dGSJ peaks with decreasing junction resistance reveals that the fractional edge states lie in the weak-coupling regime and $J_K<<\Delta$, as considered in the manuscript.

\begin{figure*}[h]
        \begin{center}
			\includegraphics[width=\textwidth]{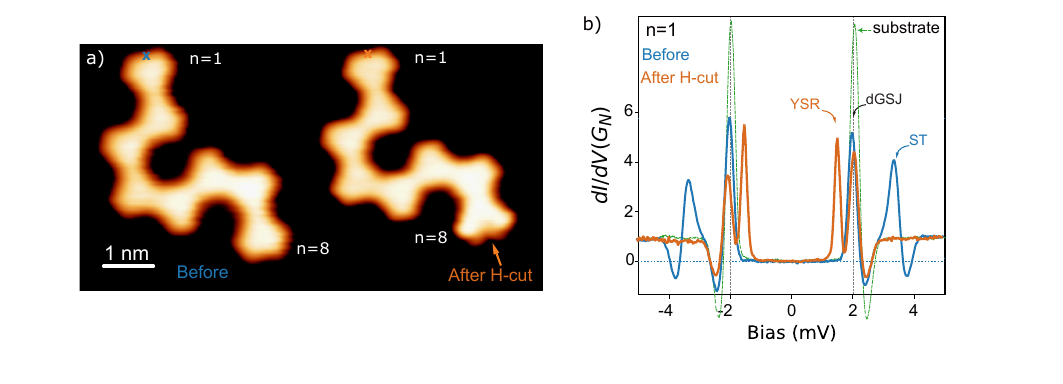}    
        \end{center}
			\caption{\textbf{Edge state passivation by dehydrogenation of a terminal triangulene - N=8. }
            a) Constant current STM image of a $N$=8 chain before and after dehydrogenation of the lower triangulene end. b) Comparison of spectra on the top triangulene before and after the dehydrogenation in a). The passivation of the lower edge state leads to disappearance of singlet-triplet (ST) features and the emergence of a in-gap YSR state. Removing the H atom quenches the intra-chain exchange correlations, leaving a free-like edge spin at the opposite end.
   }
		\label{deH2}
		\end{figure*}

\newpage

\clearpage
\section*{Complementary theoretical simulations}\label{SI:model}
\subsection{Spin model of for the S=1 chain}

\begin{figure}[h]
    \centering
    \includegraphics[width=0.98\linewidth]{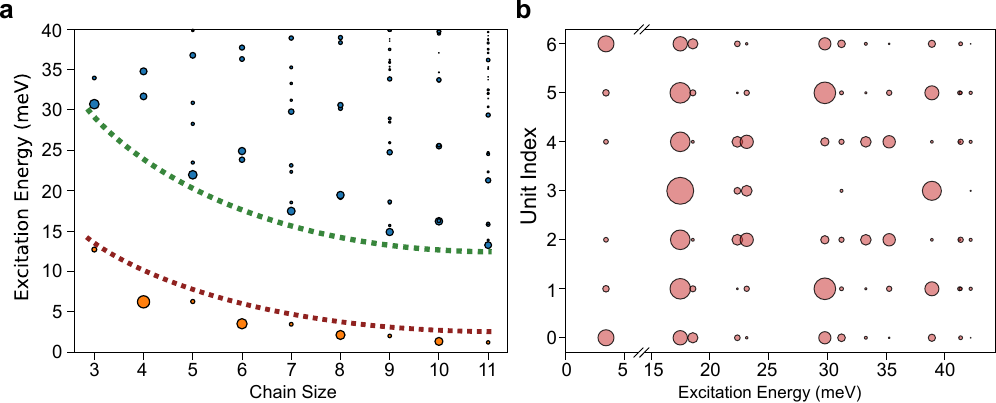}
    \caption{\textbf{Theoretical calculation of the spin excitations using the BLBQ model.} a) Spin excitations with energy less than 40~meV as a function of chain length $N$. The size of the dot represents their relative transition probability. The red and green dashed lines mark the exponential correlation decay of the edge excitations and Haldane gap, respectively. b) Site dependent distribution of spin excitation in a N=7 chain. The size of the dot represents the transition probability.}
    \label{fig:sup5}
\end{figure}

As shown in Ref.~\cite{Mishra2021}, the spin-1 triangulene chains are well described with a Bilinear Biquadratic (BLBQ) Hamiltonian,
\begin{equation}\label{Hamiltonian_chain}
    \hat H=\sum_{\langle ij\rangle}J_0\left(\vec{S}_i\cdot\vec{S}_j+\beta \left(\vec{S}_i\cdot\vec{S}_j \right)^2 \right),
\end{equation}
where the sum is taken for nearest neighbours, with $J_0\sim18$~meV and $\beta\sim 0.09$~\cite{Mishra2021}.
Supplementary Figure~\ref{fig:sup5}a shows the excitation energies from the ground state to all states below $40$~meV obtained by diagonalizing Eq.~\eqref{Hamiltonian_chain}. 

Two types of excitations can be distinguished in SFig.~\ref{fig:sup5}a, indicated by different colours: the collective bulk excitations (green dots, CS in the text), the lower of which is related to the Haldane gap~\cite{Haldane1983} in the infinite chain limit, and the edge excitations, related to exchange coupling $J$ between the fractional spin-1/2 edge excitations through the bulk Haldane gap (orange dots, ST or TS in the manuscript, for singlet-triplet or triplet-singlet, respectively). 

\textbf{Parity dependent excitation cross-section}: The area of the circles in Fig.~\ref{fig:sup5}a, and stars in Fig. 4d of the main manuscript,  is proportional to the transition probability from the ground state to the higher energy state $\ket{m}$, calculated as 
\begin{equation}\label{eq:CSec}
    \text{Prob.}=\frac{1}{N}\sum_{a,n,i}P_i|\bra{m}S^a_n\ket{i}|^2
\end{equation} 
with $a=x,y,z$, $n\in[1,N]$ and $i$ is the ground state. Note that we are averaging over all sites of the N-length chain. We also considered a possible degenerate ground state by adding the Boltzmann distribution $P_i$. 

At $T = 1.3$~K, corresponding to the experimental temperature, either the non-degenerate singlet state (even-length chains) or the triplet manifold (odd-length chains) is populated in the edge-interacting limit considered here. For an isolated chain, the transitions between singlet and triplet states are equivalent, independent to the ground state configuration. However, the singlet–triplet (ST) transition appears three times more intense than the reverse (TS) transition (larger circles in SFig.~\ref{fig:sup5}a and Fig.~4d in the main text). This is a consequence of the threefold degeneracy of the triplet state when it is the ground state. The thermal population is equally distributed among the three triplet components, so the total transition probability is shared among them, reducing the intensity of TS excitations by a factor of three. Accordingly, in the experiment, TS peaks are always much smaller than ST. Furthermore, TS peaks are frequently below one third of size of ST peaks. This larger difference could be attributed to the effect of a stronger interaction $J^*_K$ of the triplet states with the substrate. As shown elsewhere, the Kondo coupling with a substrate renormalizes spin excitation energy \cite{Oberg2014,zitko2008} and cross-section \cite{Kezilebieke2019b,trivini2023cooper}. 
 
\textbf{Distribution of excitation cross-section along the chains}: Owing to the small size of the chains, the collective spin modes appear as \textcolor{black}{many-body spin excitation} standing waves, mapped out in Fig.~4a and 4b in the manuscript. Using equation \eqref{eq:CSec},  we computed the local transition probability in each triangulene unit, obtaining distributions of excitation cross-section as shown in SFig~\ref{fig:sup5}b for a N=7 length chain. The lowest lying excitations, related to the triplet-singlet transition, are localized at the chain's   termini. On the other hand, the lowest collective \textcolor{black}{many-body spin} excitation is centred over the middle of the chain. The number of nodes increases for higher excitations, as shown in Figs. 4b and 4c for 11 and N=8 chains, respectively,  and in Supplementary Fig. 5 for N=5.

\subsection{Two-site model of YSR excitations in fractional spin end states}\label{single-site}

The interaction of a magnetic impurity with a superconducting substrate gives rise to in-gap bound states known as Yu–Shiba–Rusinov (YSR) states~\cite{yu1965bound,shiba1968classical,rusinov1969}. For a classical localized spin, the energy of these excitations is given by
\begin{equation}
    E_{YSR}=\pm \Delta \dfrac{1-\alpha^2}{1+\alpha^2},
\end{equation}
where $\alpha=\pi \nu_0 J_K S$, with $J_K$ being the exchange coupling between the impurity and the substrate, $S$ the impurity spin, and $\nu_0$ the normal-state density of states at the Fermi level.

When more than one impurity is present, the spatial overlap of the YSR wavefunctions leads to the formation of bonding and antibonding YSR states. This physics can be captured within the classical approximation~\cite{Ortuzar2022,Trivini2024}, which successfully describes the distortion of individual YSR states arising from the hybridization with neighbouring impurities. In this framework, the overlap between YSR wavefunctions provides an intuitive description of the resulting in-gap spectrum.

However, the classical approximation neglects the quantum nature of the impurity spins and therefore cannot account for their internal spin dynamics. When such dynamics become relevant~\cite{trivini2023cooper}, it is necessary to go beyond this approximation. The most accurate, albeit computationally demanding, approach to treat quantum spins is the numerical renormalization group (NRG) theory~\cite{Huang2026}.

In the following, we employ a simplified \textit{single-site Hamiltonian}~\cite{Oppen2021,Schmid2022} that explicitly incorporates the quantum nature of the impurity spin and allows for a qualitative description of YSR states induced by localized spins at edge states interacting with the superconducting substrate with strength $J_K$. In particular, the model describes the influence of inter-edge exchange coupling $J$ on the YSR bound state energy as a renormalized Kondo coupling strength $J^*_K$.  This effective model has proven successful in reproducing experimental observations~\cite{trivini2023cooper,liebhaber2022} and provides a reliable qualitative approximation for magnetic impurities in superconductors~\cite{Huang2026,Ortuzar2023}. The results of this model are used to explain the parity-dependent coupling of edge states to the superconductors, shown in Fig. 3c and 3d. 

\begin{figure}[h]
    \centering
    \includegraphics[width=0.58\linewidth]{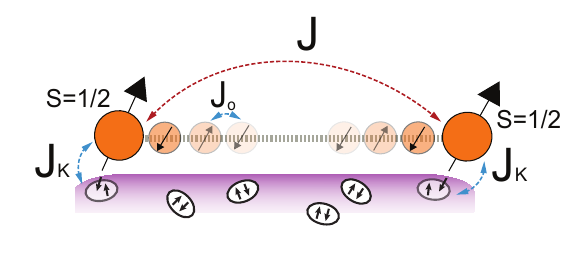}
    \caption{Schematic of a spin chain and the various exchange coupling constants used in the theoretical modelling.}
    \label{sfig:model}
\end{figure}

At low applied bias, the inner triangulene units of a chain form a valence-bond solid~\cite{Mishra2021,Huang2026}. Since the inter-triangulene exchange $J_0$ dominates over $J_K$, the formation of local spin singlets suppresses the exchange coupling of individual triangulenes to the surface, and \textcolor{black}{many-body spin} excitations are protected from coupling to the substrate by the Haldane gap. 
In contrast, the terminal units host unpaired spin-$1/2$ edge states, which are decoupled from the bulk of the chain and can interact directly with the superconducting substrate and with each other, with strength $J_K$ and $J$, respectively. Consequently, the interaction between the chain and the substrate can be effectively modelled by two exchange-coupled spin-$1/2$ impurities and their interaction to the superconducting substrate~\cite{Huang2026}, as shown in Fig. \ref{sfig:model},
\begin{equation}
    H=\sum_{i=1,2}\left(J_{Ki}\hat{c}^{\dagger}_{i\sigma}\vec{S}_i\cdot \vec{s}_{\sigma\sigma'}\hat{c}_{i\sigma'}+\Delta\hat{c}^{\dagger}_{i\uparrow}\hat{c}^{\dagger}_{i\downarrow}+\mathrm{H.c.}\right)+J\vec{S}_1\cdot\vec{S}_2.
\end{equation}
Here, $\vec{S}_i$ denotes the spin operator of the $i$-th edge state, $J_{Ki}$ its exchange coupling to the substrate, and $J$ the effective inter-impurity exchange interaction accounting for singlet–triplet (or triplet–singlet) excitations of the edge states. The coupling $J$ is positive (antiferromagnetic) for even chains and negative (ferromagnetic) for odd chains. We neglect any hopping between superconducting sites due to the large separation between the edges, i.e., we assume that the direct exchange between edge states dominates over any RKKY-type interaction.

This effective Hamiltonian captures the low-energy excitations of the triangulene chain, with all internal degrees of freedom of the inner units integrated out. In what follows, we assume that the exchange coupling to the substrate is smaller than $\Delta$ ($ J_{Ki} << \Delta  $), such that the superconductor remains in the weak-coupling regime \cite{Franke11,hatter_scaling_2017}. In this case, all YSR excitations with energies $E<\Delta$ change the fermion parity of the superconducting system from even (ground state) to odd (excited state), as discussed in Refs.~\cite{Oppen2021,Ortuzar2023,trivini2023cooper}. The ground-state and excited-state energies are then given by

\begin{align} \label{eq:s8}
    &E_{GS} = \begin{cases}
        -2\Delta - \dfrac{3}{4}J & \text{if } J > 0, \\[6pt]
        -2\Delta + \dfrac{1}{4}J & \text{if } J < 0,
    \end{cases} \\[6pt]
    &E_i = \dfrac{1}{4}\left(-J - J_{Ki} - 2\sqrt{J^2 - J\,J_{Ki} + J_{Ki}^2} - 4\Delta\right).
\end{align}

where $E_i$ corresponds to a single-electron excitation localized at the $i$-th edge. In the limit $J=0$, these expressions reduce to the single-impurity YSR energies. By plotting $E_i(J)-E_{\mathrm{GS}}(J)$ as a function of $J$  one can qualitatively capture the renormalization of the YSR excitation energies induced by the inter-impurity exchange coupling, as shown in Fig.~3d of the main text. In the non-interacting limit expected for sufficiently long chains, ${J\rightarrow0}$, these expressions describe the YSR excitation of independent spin-1/2 impurities,  $ (E_i(0)-E_{\mathrm{GS}}(0))=\Delta-3J_K/4$.

For symmetric impurity couplings ($J_{K1}=J_{K2}=J_K$), two distinct behaviours emerge depending on the sign of $J$ \eqref{eq:s8}. For $J>0$, the ground state is a singlet formed between the two impurity spins. As $J$ increases, the independent spin-$1/2$ character of the impurities is lost, and the system effectively behaves as a non-magnetic impurity. The energy difference $\lim_{J\rightarrow \infty} 
(E_i(J)-E_{\mathrm{GS}}(J))=\Delta$, 
yielding a vanishing effective Kondo coupling, $J^*_{\mathrm{K}}\rightarrow0$~\cite{zitko2011}, as depicted in Fig. 3c.

Conversely, for $J<0$, the ground state is a triplet formed by the two impurity spins. As $|J|$ increases, the impurities again lose their independent character, but the system instead realizes an effective spin-$1$ impurity coupled symmetrically to two YSR screening channels, corresponding to a two-channel Kondo scenario~\cite{jones1987}. In this case, the energy difference 
$\lim_{J\rightarrow \infty} (E_i(J)-E_{\mathrm{GS}}(J))=\Delta-J_K/2$. Comparing it to the single spin-1/2 impurity case~\cite{Oppen2021} the effective coupling satisfies $J^*_{\mathrm{K}}/J_K\rightarrow 2/3$ (Dashed line in Fig. 3c), reflecting that only the triplet states with finite $S_z$ projection couple to the substrate. The validity of this two-site model has been benchmarked against full NRG calculations in Refs.~\cite{Oppen2021,Huang2026}. 

This simple model qualitatively describes the low-energy spectrum and the characteristics of the many-body ground state. It is interesting to note that our experiments only show YSR states in the weak-coupling regime (see Fig. S15), where the ground state is fully described by the relative alignment of the two quantum spins, i.e., the system is either in a singlet or a triplet state. 
This description is expected to break down at larger values of Kondo coupling $J_K$, when the entanglement with the superconducting sites become as important as that between the two edge spins. As shown in Ref.~\cite{Huang2026}, this can give rise to more complex ground states built from Kondo singlets with the substrate and their interaction through the chain. We expect such physics to be observable in longer chains, where the exchange coupling $J$ becomes comparable $J_K$.

\bibliographystyle{apsrev4-1} 

%